\documentclass[12pt]{article}

\usepackage{latexsym}
\usepackage{amssymb,amsfonts,amsmath}
\usepackage{graphicx} 
\usepackage{indentfirst}
\usepackage{bbm}
\usepackage{amssymb}
\usepackage{verbatim}
\usepackage{amsmath, amsthm,amssymb}
\usepackage{mathrsfs}
\usepackage{hyperref}
\usepackage{amsfonts}
\usepackage{dsfont}
\usepackage{cite}

\parskip=\medskipamount

\newcommand {\cN}{{\cal N}}

\def\a{\alpha}

\def\b{\beta}

\def\d{\delta}

\def\l{\lambda}

\def\n{\nu}

\def\rd{{\rm d}}
\def\ri{{\rm i}}
\def\re{{\rm e}}

\newcommand{\ad}{{\dot{\alpha}}}                           
\newcommand{\bd}{{\dot{\beta}}}                            
\newcommand{\ve}{\varepsilon}                            

\newcommand{\pa}{\partial}                           
\newcommand{\hf}{\frac12}

\newcommand{\vf}{\varphi}

\newcommand{\be}{\begin{equation}}
\newcommand{\ee}{\end{equation}}
\newcommand{\bea}{\begin{eqnarray}}
\newcommand{\eea}{\end{eqnarray}}

\def\double #1{#1{\hbox{\kern-2pt $#1$}}}

\newif\ifdtup

\newcommand{\bsubeq}{\begin{subequations}}
\newcommand{\esubeq}{\end{subequations}}

\numberwithin{equation}{section}

\begin{document}

\begin{titlepage}
\begin{flushright}
August, 2019 \\
\end{flushright}
\vspace{5mm}

\begin{center}
{\Large \bf 
Manifestly duality-invariant interactions in diverse dimensions}
\end{center}

\begin{center}

{\bf Sergei M. Kuzenko} \\
\vspace{5mm}

\footnotesize{
{\it Department of Physics M013, The University of Western Australia\\
35 Stirling Highway, Perth W.A. 6009, Australia}}  
~\\

\vspace{2mm}

\end{center}

\begin{abstract}
\baselineskip=14pt
As an extension of the Ivanov-Zupnik approach to self-dual nonlinear electrodynamics
in four dimensions \cite{IZ1,IZ2}, we reformulate  U(1) 
duality-invariant nonlinear models for a gauge $(2p-1)$-form in $d=4p$ dimensions 
as field theories with manifestly U(1) invariant self-interactions. This reformulation 
is suitable to generate arbitrary duality-invariant nonlinear systems including those with higher derivatives.
\end{abstract}
\vspace{5mm}

\vfill

\vfill
\end{titlepage}

\newpage
\renewcommand{\thefootnote}{\arabic{footnote}}
\setcounter{footnote}{0}


\allowdisplaybreaks


\section{Introduction}

As an extension of the seminal work by Gaillard and Zumino \cite{GZ1}, 
the general formalism of duality-invariant
models for nonlinear electrodynamics in four dimensions was developed in the mid-1990s \cite{GR1,GR2,GZ2,GZ3}. The Gaillard-Zumino-Gibbons-Rasheed  
(GZGR) approach was generalised to off-shell $\cN=1$ and $\cN=2$ globally \cite{KT1,KT2} and locally \cite{KMcC,K12} supersymmetric theories. 
In particular, the first consistent 
perturbative scheme to construct the $\cN=2$ supersymmetric Born-Infeld action 
was given in \cite{KT2} (this approach was further pursued in \cite{BCFKR}). 
The GZGR formalism was also extended to higher dimensions
\cite{Tanii,AT,ABBZ}.

Nonlinear electrodynamics with U(1) duality symmetry is described by a 
 Lorentz invariant Lagrangian $L(F_{ab})$ which is a solution 
to the self-duality equation 
\bea
\widetilde{G}^{ab}G_{ab}  +  \widetilde{F}^{ab}F_{ab} = 0~,
\label{1.1}
\eea
where
\bea
\widetilde{G}^{ab} (F):=
\hf \, \ve^{abcd}\, G_{cd}(F) =
2 \, \frac{\pa L(F)}{\pa F_{ab}}~.
\eea
In the case of theories with higher derivatives, this scheme is generalised in accordance with the two rules given in \cite{KT1}. 
Firstly, the definition of $\widetilde G$ is replaced with
\bea
\widetilde G^{ab}[F] =2 \,{\delta S[F] \over\delta F_{ab}}~.
\eea
Secondly, the self-duality equation \eqref{1.1} is replaced with 
\bea
\int \rd^4 x \Big( \widetilde{G}^{ab}G_{ab}  +  \widetilde{F}^{ab}F_{ab} \Big)= 0~.
\eea
Duality-invariant theories with higher derivative theories naturally occur in 
$\cN=2$ supersymmetry \cite{KT2}.
Further aspects of duality-invariant theories 
with higher derivatives were studied in, e.g., \cite{AFZ,Chemissany:2011yv,AF,AFT}.

Self-duality equation \eqref{1.1} is nonlinear, 
and therefore its general solutions are difficult to find.
 In the early 2000s, Ivanov and Zupnik \cite{IZ1,IZ2}
proposed a reformulation of duality-invariant electrodynamics
 involving  an auxiliary antisymmetric tensor $V_{ab}$,  
which is  equivalent to a symmetric spinor
$V_{\a\b}= V_{\b\a}$ and its conjugate $\bar  V_{\ad\bd}$.\footnote{This approach 
was inspired by the structure on the $\cN=3$ supersymmetric Born-Infeld action
proposed in \cite{IZ_N3}.}
 The new Lagrangian $L(F,V)$ is at most quadratic in
the electromagnetic field strength $F_{ab}$, while the self-interaction is described 
by a nonlinear function of the auxiliary variables, $L_{\rm int} (V_{ab})$,
\bea
 {L}(F_{ab} , V_{ab}) = \frac{1}{4} F^{ab}F_{ab} +\hf V^{ab}V_{ab} 
 - V^{ab}F_{ab} + L_{\rm int} (V_{ab})~.
\eea 
The original theory  $L(F_{ab})$ is obtained from 
${L}(F_{ab} , V_{ab})$ by integrating out the auxiliary variables. 
In terms of ${L}(F_{ab} , V_{ab})$, the condition of U(1) duality invariance 
was shown \cite{IZ1,IZ2}  to be equivalent to the requirement that the self-interaction
\bea
L_{\rm int} (V_{ab}) = L_{\rm int} (\n, \bar \n)~, \qquad \n:=V^{\a\b}V_{\a\b}
\eea
is invariant under linear U(1)  transformations $\n \to \re^{\ri \vf} \n$, with $\vf \in \mathbb R$,
and thus 
\bea
L_{\rm int} (\n, \bar \n)= f (\n \bar \n)~,
\label{1.5}
\eea
where $f$ is a real function of one real variable. 
The Ivanov-Zupnik (IZ) approach \cite{IZ1,IZ2} has been used by 
Novotn\'y \cite{Novotny} to establish the relation between helicity conservation for the tree-level scattering amplitudes and the electric-magnetic duality.

The above discussion shows that the IZ approach is a universal formalism to generate  U(1) duality-invariant models for nonlinear electrodynamics. 
Some time ago, there was a revival of interest in duality-invariant dynamical systems
 \cite{BN,CKR,Chemissany:2011yv} inspired by the desire to achieve a better understanding
of   the UV properties of extended supergravity theories. 
The authors of \cite{BN} 
have put forward the so-called ``twisted self-duality constraint,''
which was further advocated in  \cite{CKR,Chemissany:2011yv}, 
as a systematic procedure to generate manifestly duality-invariant theories.
However, these approaches  have been demonstrated \cite{IZ3}
to be variants of the IZ scheme \cite{IZ1,IZ2} developed a decade earlier. 

The IZ approach has been generalised to off-shell $\cN=1$ and $\cN=2$ globally  and locally supersymmetric theories \cite{K13,ILZ}. In this note we provide a generalisation 
of the approach to higher dimensions, $d=4p$.
In even dimensions, $d=2n$, 
the maximal duality group for a system of $k$ gauge $(n-1)$-forms depends
on the dimension of spacetime. The duality group is U$(k)$ if $n$ is even, 
and ${\rm O}(k) \times {\rm O}(k) $ if $n $ is odd \cite{AT}
(see, e.g, section 8 of \cite{KT2} for a review).
This is why we choose $d=4p$.
The fact that the maximal duality group
depends on the dimension of space-time
was discussed in the mid-1980s \cite{Ta2,CFG} and also in 
the late 1990s \cite{CJLP1,CJLP2}.


\section{New formulation}

In Minkowski space of even dimension $d=4p \equiv 2n$, with $p$ a positive integer,
we  consider
a self-interacting theory of a  gauge $(n-1)$-form $A_{a_1 \dots a_{n-1}}$
with the property that the Lagrangian, $L=L (F)$,
 is a function of the field strengths 
$F_{a_1 \dots a_n} =n  \pa_{ [a_1} A_{a_2 \dots a_{n} ] }$.\footnote{We follow the notation and conventions of \cite{KT2}.}
We assume that the theory possesses U(1) duality invariance. 
This means that the Lagrangian is a solution
to the self-duality equation \cite{AT}
\bea
\widetilde G^{a_1 \dots a_n} G_{a_1 \dots a_n} 
+ \widetilde F^{a_1 \dots a_n} F_{a_1 \dots a_n} =0~,
\label{self-duality}
\eea
where we have introduced
\bea
\widetilde{G}^{a_1 \dots a_n} (F)= n!
\frac{\pa L (F)}{\pa F_{ a_1 \dots a_n}}~.
\eea
As usual, the notation $\widetilde F$ is used  for the Hodge dual of $F$, 
\bea
\widetilde{F}^{a_1 \dots a_n} =
\frac{1}{n!} \, \ve^{a_1 \dots a_n b_1 \dots b_{n}} \,
F_{b_1 \dots b_{n} }~.
\eea

We now introduce a reformulation of the above theory. 
Along with the field strength $F_{a_1\dots a_n}$, our new Lagrangian 
$L(F, V)$ is defined to depend on an auxiliary rank-$n$ 
antisymmetric tensor $V_{a_1 \dots a_n}$ which is unconstrained. We choose $L(F, V)$ to have the form
\bea
L(F,V) = \frac{1}{n!} \Big\{ \hf F \cdot F +  V \cdot V - 2 V \cdot F\Big\} 
+ L_{\rm int} (V) ~,
\label{first-order}
\eea
where we have denoted
\bea
 V\cdot F:= V^{a_1 \dots a_n} F_{a_1 \dots a_n}~.
 \eea
 The last term in \eqref{first-order},  $L_{\rm int} (V) $, is at least quartic 
 in $V_{a_1 \dots a_n}$.
It is assumed that the equation of motion for $V$, 
\bea
\frac{\pa}{\pa V^{a_1 \dots a_n} } L(F,V) =0~,
\eea
allows one to integrate out the auxiliary field $V$ to result with $L(F)$. 

It may be shown that the self-duality equation \eqref{self-duality} 
is equivalent to  the following condition on the self-interaction in \eqref{first-order}
\bea
\widetilde V^{a_1 \dots a_n} \frac{\pa}{\pa V^{a_1 \dots a_n} } L_{\rm int} (V) =0~.
\label{2.7}
\eea
Introducing (anti) self-dual components of $V$, 
\bea
V_\pm^{a_1\dots a_n} = \hf \Big( V^{a_1\dots a_n}  
\pm \ri \widetilde V^{a_1\dots a_n} \Big) ~, \qquad 
\widetilde V_\pm = \mp\ri V_\pm ~,\qquad V = V_+ +V_-~,
\eea
the above condition turns into 
\bea
\Big( 
 V_+^{a_1 \dots a_n} \frac{\pa}{\pa V_+^{a_1 \dots a_n} } 
 - V_-^{a_1 \dots a_n} \frac{\pa}{\pa V_-^{a_1 \dots a_n} } \Big)L_{\rm int} (V_+, V_-) 
=0~.
 \eea
This means that $ L_{\rm int} (V_+, V_-) $ is invariant under U(1) phase transformations, 
\bea
L_{\rm int} (\re^{\ri \vf}  V_+, \re^{-\ri \vf} V_-)  = L_{\rm int} (V_+, V_-) ~, \qquad 
\vf \in {\mathbb R}~.
\eea
In four dimensions, the most general solution to this condition is given by eq. 
\eqref{1.5}. Similar solutions exist in higher dimensions, 
$ L_{\rm int} (V_+, V_-) = f (V_+ \cdot V_+ V_- \cdot V_-)$, with $f(x)$ a real function 
of one variable. However more general 
self-interactions become possible beyond four dimensions.

It is worth pointing out that an infinitesimal U(1) duality transformation 
\begin{subequations}
\bea
\d   \left( \begin{array}{c}  G \\ F \end{array} \right)
=  \left( \begin{array}{cr} 0 & 
- \l \\  \l  &  0 \end{array} \right) 
\left( \begin{array}{c}  G \\ F  \end{array} \right) 
\eea
leads to the following transformation of $V$
\bea
\d V = \l \widetilde V~.
\eea
\end{subequations}
Equation \eqref{2.7} tells us that $ L_{\rm int} (V) $ duality invariant. 

There are several interesting generalisations of the construction described. 
They include (i) coupling to gravity; (ii) coupling to a dilaton with enhanced 
SL$(2,{\mathbb R} )$ duality; (iii) duality-invariant systems with higher derivatives;
 and (iv) U$(k)$ duality-invariant systems of $k$ gauge $(2p-1)$-forms in $d=4p$ dimensions.

Recently, U(1) duality-invariant theories  of a gauge $(2p-1)$-form in $d=4p$ dimensions 
have been described \cite{Buratti:2019cbm}
within the Pasti-Sorokin-Tonin approach \cite{PST1,PST2}.
It was argued in \cite{Buratti:2019cbm} that the approach of \cite{PST1,PST2}
is the most efficient method to determine all possible manifestly U(1) 
duality invariant self-interactions provided Lorentz invariance is kept manifest.
Our analysis has provided an alternative formalism.\footnote{It is worth pointing out that
in four dimensions a hybrid formulation  has been developed \cite{INZ} which combines the powerful features
of the IZ approach with thoes advocated in \cite{PST1,PST2}.}
\\


\noindent
{\bf Acknowledgements:}\\
I am grateful to Stefan Theisen for useful comments.
This work is supported in part by the Australian 
Research Council, project No. DP160103633.

\begin{footnotesize}

\end{footnotesize}


\end{document}